# Parametrically driven inertial sensing in chip-scale optomechanical cavities at the thermodynamical limits with extended dynamic range

Jaime Gonzalo Flor Flores[1], Talha Yerebakan[1], Wenting Wang[1], Mingbin Yu[2], Dim-Lee Kwong[2], Andrey Matsko[3], and Chee Wei Wong[1]

[1] Fang Lu Mesoscopic Optics and Quantum Electronics Laboratory, University of California, Los Angeles, CA 90095, USA

[2] Institute of Microelectronics, A*STAR, Singapore 117865, Singapore

[3] Jet Propulsion Laboratory (JPL), California Institute of Technology, Pasadena, CA 91109, USA
E-mail: jflorflores@ucla.edu; cheewei.wong@ucla.edu

**Recent scientific and technological advances have enabled the detection of gravitational waves** [1]**, autonomous driving** [2]**, and the proposal of a communications network on the Moon (Lunar Internet or LunaNet)** [3]**. These efforts are based on the measurement of minute displacements and correspondingly the forces or fields transduction, which translate to acceleration, velocity, and position determination for navigation. State-of-the-art accelerometers use capacitive or piezo resistive techniques** [4, 5, 6]**, and micro-electromechanical systems (MEMS)** [7] **via integrated circuit (IC) technologies in order to drive the transducer and convert its output for electric readout. In recent years, laser optomechanical transduction** [8, 9] **and readout have enabled highly sensitive detection of motional displacement. Here we further examine the theoretical framework for the novel** **mechanical frequency readout technique of optomechanical transduction when the sensor is driven into oscillation mode** [8]**. We demonstrate theoretical and physical agreement and characterize the most relevant performance parameters with a device with 1.5mg/Hz acceleration sensitivity, a 2.5 fm/Hz$^{1/2}$ displacement resolution corresponding to a 17.02 μg/Hz$^{1/2}$ force-equivalent acceleration, and a 5.91 Hz/nW power sensitivity, at the thermodynamical limits. In addition, we present a novel technique for dynamic range extension while maintaining the precision sensing sensitivity. Our inertial accelerometer is integrated on-chip, and enabled for packaging, with a laser-detuning-enabled approach.**

## I. Driving and readout circuit components for resonant accelerometers

Resonant accelerometers operate by driving a mechanical structure (e.g., MEMS) with an electrical circuit and measuring the changes in frequency due to input acceleration. By applying feedback to a resonant accelerometer, creating an oscillator, the resulting high-quality factor ($Q_m$) is represented as a reduction on phase noise and a better bias stability. Such increase in performance comes with an increase in complexity on the electrical circuit necessary for its operation. At the same time, noise components introduced by the circuit limit the performance of the entire system.

Multiple architectures can be used for the readout circuit on an oscillating accelerometer [10]. **Figure 1** shows a standard design that illustrates the multiple components required for the readout and driving circuits. First, the resonant transducer is connected to a front-end amplifier that can be a variation of a transimpedance amplifier (TIA), such as a two-stage TIA or a T-network. The function of this section is to amplify the motional current generated by the transducer and convert it into a voltage signal. A front-end amplifier needs to have a bandwidth of at least 10 times that of the resonant frequency of the transducer to avoid introducing phase shift; in addition, it needs an input current noise lower than current produced by the minimum transducer movement.

A variable gain amplifier (VGA) is connected along the feedback path, and drives the oscillator. An amplitude control module is sometimes integrated in the main loop, or added as a secondary one, in order to maintain the amplitude of the oscillations constant. In a resonant accelerometer, large mechanical displacements introduce nonlinear effects in the mechanical motion. In some devices, even smaller displacements can cause changes in the effective stiffness of the system, which can present themselves as changes in the resonant frequency, inducing noise in the measurements. An amplitude control module is composed of a buffer, an amplitude detector, an external amplitude reference signal, a filter, and an error amplifier whose inputs are the detector and reference signals [11]. A chopper could be connected after the filter in order to

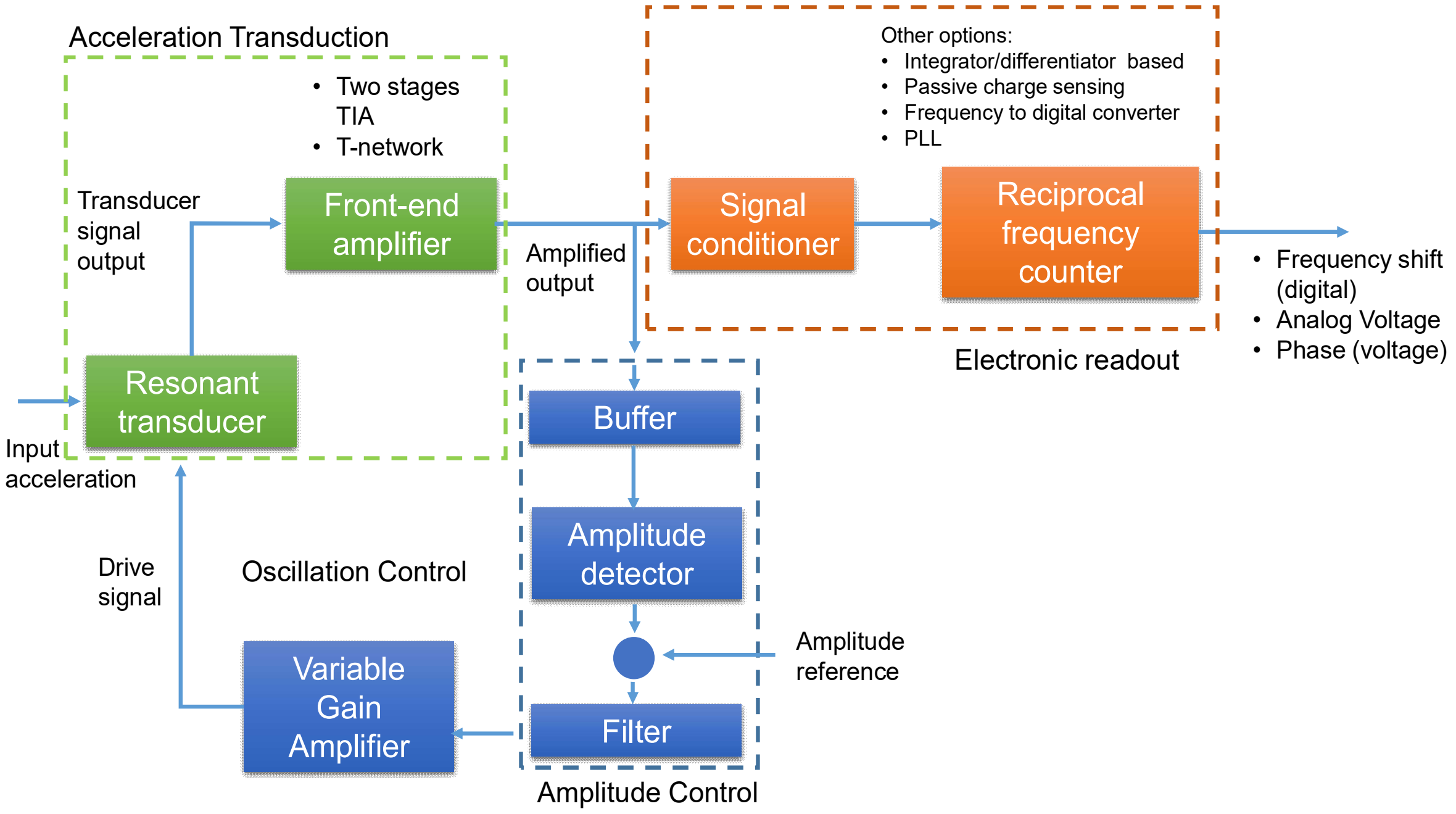

**Figure 1.- Resonant accelerometer driving and readout.**

Diagram shows the multiple electrical components necessary for the operation of a traditional state-of-the-art resonant accelerometer.

limit the oscillation amplitude to a fixed value and avoid entering into strong nonlinear regime. Finally, a frequency measurement circuit can be used for the output signal in order to improve bias-instability of the oscillating accelerometer.

As stated before, the performance of the feedback control loops is critical for the operation of the oscillating accelerometer, and its design and implementation are not trivial. However, as described in the theoretical sections of this work, the proposed optomechanical accelerometer has an integrated optical feedback system that drives the oscillator, which is an important advantage of this architecture.

## II. Optomechanical accelerometer optical and mechanical design.

The photonic crystal has been nanofabricated on a 250 nm thick, top membrane of the silicon-on-insulator (SOI) wafer. The circulars holes were etched using deep UV lithography and have a radius $r$ = 185 nm, and a lattice constant $a_p$ = 510 nm. The center row of holes of the photonic crystal has been removed and replaced by a slot cavity that has a width $s$ = 100 nm, as it can be seen in **Figure 2**. A lattice perturbation has been introduced in the center region, and the three most immediate rows of holes have been displaced by 5, 10, and 15 nm, as they get farther away from the slot cavity. The crystal has been optimized for maximum radiation pressure generation by reducing electric field leakage to the silicon material as much as possible. The cavity mode volume has been calculated at $0.051(\lambda/n)^3$. The cavity bandgap has been designed to permit three optical modes. The fundamental mode was designed at ~1550 nm, and has the strongest coupling to the fundamental mechanical mode.

The mechanical structure was fabricated on the same membrane of the SOI wafer, and consists of a stationary mass and a resonant mass that is suspended by four cantilever beams. The design of these beams has been optimized for acceleration sensing and a 41 kHz resonant frequency. The effective mass of the stationary mass is 7.2 nano-grams.

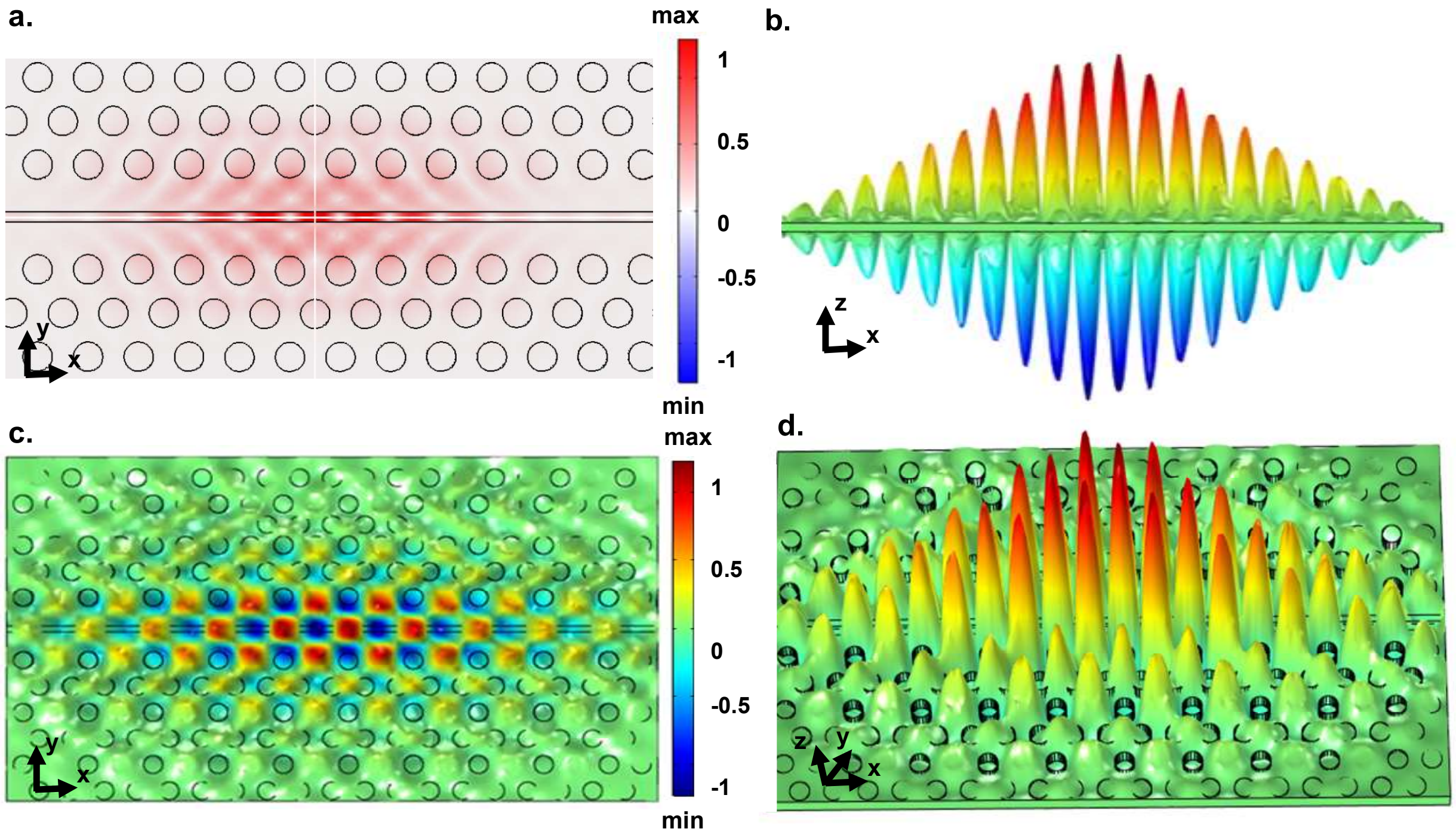

**Figure 2.- Opitcal mode design.**

**a-d** represent the calculated optical mode for the fully integrated photonic crystal and tunneling waveguide used to couple input light into the center cavity and to take the output modulated signal into the detector. Axis marks the orientation of the photonic crystal, and scale bar shows the intensity of the electric field. Sub-wavelength mode confinement obtained with a mode volumen $0.051(\lambda/n)^3$ on the central slot cavity.

## III. Traditional optical frequency shift acceleration detection techniques

Previous studies have demonstrated optomechanical transducers capable of measuring continuous forces and displacements [12], torque [13], and specific forces [14, 15]. In general, measurement of displacements can be correlated to a force or acceleration that produced it, and the standard optomechanical approach uses changes in the displacement spectra to quantify it. At the same time, changes in optical frequency shift can also be used [14], as illustrated in **Figure S3**. However, in the presented work, changes in mechanical resonance are used to measure specific force as described in the main text.

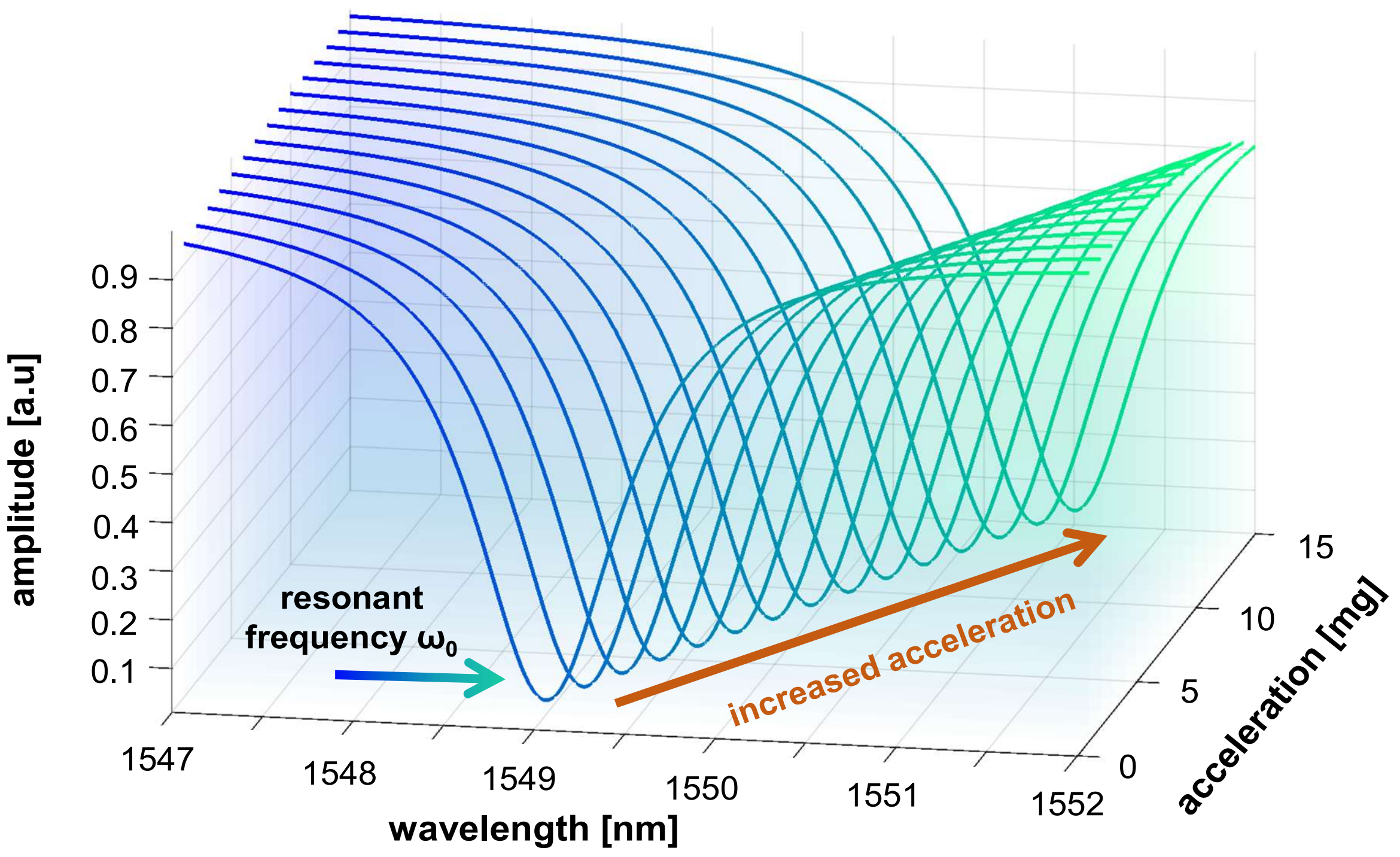

**Figure 3.- Opitcal frequency shift acceleration detection.**

Multiple traces show the optical transmission of a sample cavity as a function of laser detuning. The deep in the transmission is produced at the cavity's resonant wavelength. As multiple accelerations are applied to the cavity, the optical resonance shifts proportionally.

## IV. Optomechanical accelerometer RF readout

The presented optomechanical accelerometer is composed of two coupled oscillators. The mechanical oscillator can be modeled as a damped oscillator without loss of generality. For a single degree of freedom system, the equation characterizing the motion can be written as:

$$m\ddot{x} + b\dot{x} + kx = 0 \tag{S1}$$

Where *m* is the mass of the resonator, *b* is the damping coefficient, and *k* the spring constant. In general, this approximation assumes a linear restoring force ($-kx$) that is dependent on displacement, and a resisting force ($-b\dot{x}$) dependent on the velocity of movement. If the mechanical resonant frequency is defined as $\omega_m = (k/m)^{1/2}$, and the damping ratio as $\zeta = b/2(k\ m)^{1/2}$, the equation can be rewritten as:

$$\ddot{x} + 2\zeta\omega_m\dot{x} + \omega_m^2 x = 0 \tag{S2}$$

The damping ratio can also be defined in terms of the mechanical quality factor as $\zeta \equiv 1/Q_m$.

As the mechanical oscillator is driven by an external force, the above equation can be written as:

$$\ddot{x} + 2\zeta\omega_m\dot{x} + \omega_m^2 x = \frac{F_o}{m} + \frac{F_T}{m} + \frac{F_S}{m} \tag{S3}$$

Where $F_o$ is the optical force, $F_T$ the thermal force, and $F_s$ the external specific force, which produces a displacement $x_s = F_s/\omega^2_m m_x$.

On the other hand, the dynamics of the optomechanical field are given by [16]:

$$\frac{d\hat{a}}{dt} = i\Delta(x)\hat{a} - \left(\frac{1}{2\tau_o} + \frac{1}{2\tau_{ext}}\right)\hat{a} + i\sqrt{\frac{1}{2\tau_{ex}}}s \tag{S4}$$

where $|\hat{a}|^2$ is the stored cavity energy and $1/(2\tau_0)+1/(2\tau_{ex})=1/(2\tau)$ is the optical field decay rate.

Solving equations S3 and S4 in terms of Bessel functions, the solution can be written as:

$$\hat{a}(t)e^{-i\omega_l t} = \sqrt{\frac{n_C}{\tau}}s\sum_{k=-\infty}^{+\infty}\frac{(-i)^k J_k(\beta)}{-i(\omega_l+k\omega_m-\omega_c+g_{om}x_s)+\frac{1}{2\tau}}e^{-i(\omega_l+k\omega_m)t+i\beta cos(\omega_m t)} \quad \text{(S5)}$$

where $\beta = g_{om}x_o/\omega_m$. Equation S5 can be further simplified into equation 3 in the case where n the case where $\omega_l$ - $\omega_c + g_{om}x_s >> \omega_m$.

## V. Optomechanical coupling rate determination

The optomechanical coupling coefficient has been defined as $g_{om}/2\pi = \mathrm{d}\omega_o/\mathrm{d}x = g_o/x_{zpf}$, where $g_o$ is the optomechanical coupling strength and $x_{zpf} = \sqrt{\frac{\hbar}{2m_{eff}\omega_m}}$ is the mechanical zero point fluctuation amplitude. In order to determine $g_{om}/2\pi$, we use a calibration tone generated by a phase modulator and compare the tone with the optomechanically induced mode [17, 18].

The Power Spectral Density (PSD) of the optomechanical mode and the calibration tone can be described as:

$$S_{\omega_m}(\omega) \approx \frac{8g_o^2 n_{th}\omega_m^2}{(\omega^2-\omega_m^2)^2+\Gamma_m^2\omega_m^2} \quad \text{(S6)}$$

$$S_{\omega_{tone}}(\omega) = \frac{1}{2}\omega_{tone}^2\beta^2\delta(\omega-\omega_{tone}) \quad \text{(S7)}$$

Where $\omega_m$ is the fundamental frequency of the mechanical mode, $\omega_{tone}$ is the modulation frequency, $\Gamma_m$ is the mechanical mode damping rate, $n_{th}$ is the phonon number at the given bath temperature, and can be estimated as the thermal energy divided by single phonon energy $k_BT/\hbar\omega_m$. Finally, $\beta$ is the modulation depth defined as $V_{tone}\pi/V_\pi$; where $V_{tone}$ is the applied voltage to the modulator and $V_\pi$ is the half-wave voltage of the modulator. The PSD of the mechanical mode and the calibration tone is further modified by a frequency-dependent transduction factor of the photodetector $G_V(\omega)$, and the final modes are given by:

$$S_{V_{\omega_m}}(\omega) = |G_V(\omega_m)|^2 S_{\omega_m}(\omega) \tag{S8}$$

$$S_{V_{\omega_{tone}}}(\omega) = |G_V(\omega_{tone})|^2 S_{\omega_{tone}}(\omega) \tag{S9}$$

When these functions are integrated in the frequency domain, the following equations are obtained:

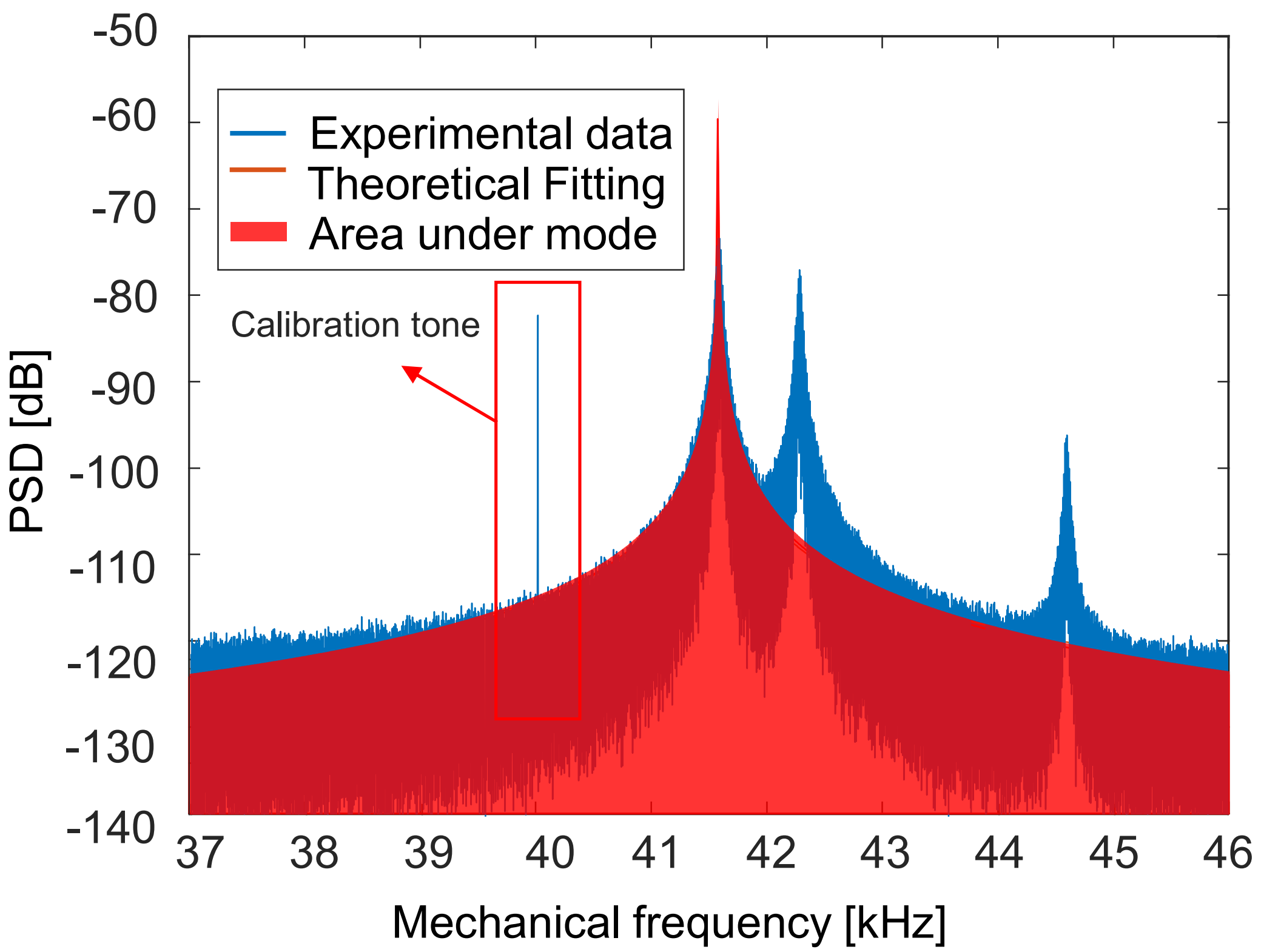

**Figure 4.- Optomechanical coupling rate determination.**

Determination of the optomechanical coupling rate by using a calibration tone. Figure shows the meassure optomechanical spectrum and the lorenztian fitting. The area under the fitting has been calculated to determine $g_{om}$.

$$A_m = |G_V(\omega_m)|^2 \int_{-\infty}^{\infty} S_{\omega_m}(\omega)\, d\omega = 2|G_V(\omega_m)|^2 g_o^2 n_{th} \quad \text{(S10)}$$

$$A_{tone} = |G_V(\omega_{tone})|^2 \int_{-\infty}^{\infty} S_{\omega_{tone}}(\omega)\, d\omega = \frac{1}{2}|G_V(\omega_m)|^2 \omega_{tone}^2 \beta^2 \quad \text{(S11)}$$

By solving equations S10 and S11, the final form for $g_o$ can be obtained as:

$$g_o = \frac{\beta\omega_{tone}}{2}\left(\frac{A_m}{n_{th}A_{tone}}\right)^{\frac{1}{2}} \left|\frac{G_V(\omega_{tone})}{G_V(\omega_m)}\right| \quad \text{(S12)}$$

Assuming that the calibration tone frequency and the center frequency of the mechanical mode are close to each other, the transduction factor can be simplified [17, 18], and equation S12 can be reduced to:

$$g_o = \frac{\beta\omega_{tone}}{2}\left(\frac{A_m}{n_{th}A_{tone}}\right)^{\frac{1}{2}} \quad \text{(S13)}$$

**Figure 4** shows the Lorentzian fit of the mechanical mode and the calibration tone. By using equation S13, the optomechanical coupling strength has been calculated to $g_o \approx 464$ kHz, and the optomechanical coupling rate $g_{om}/2\pi \approx 87.74$ GHz/nm.

## VI. Effective frequency modeling

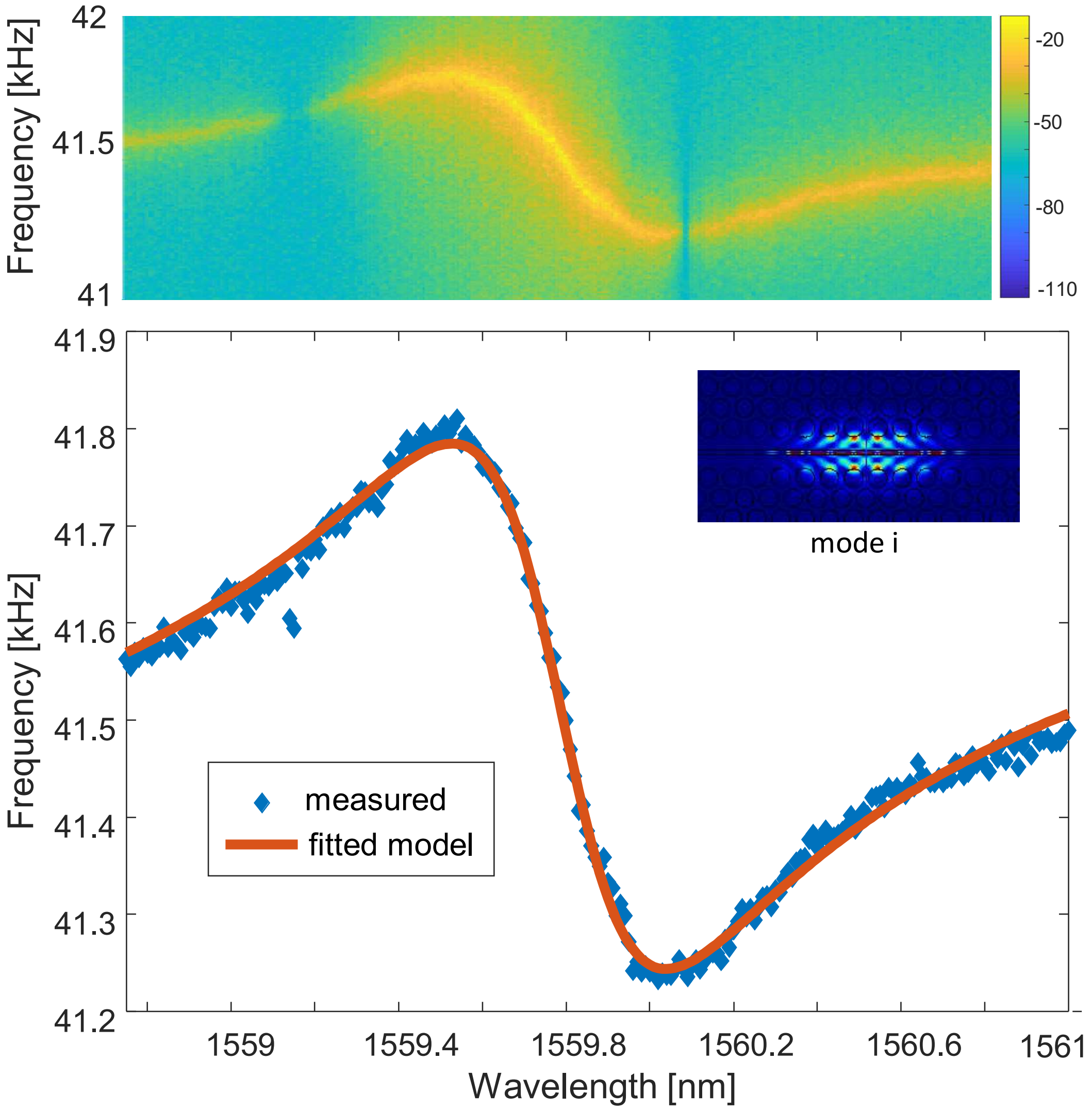

**Figure 5.- Effective mechanical frequency fitting.**

**a.** Mechanical resonant frequency as a function of driving laser wavelength for the optomechanical cavity. The color scale represents the power of the signal measured on the ESA. **b.** The effective mechanical frequency has been calculated by using equation 1 in the main text, and fitted to the resonant frequency measured from the optomechanical transducer. Inset shows a simulation of the fundamental optical mode responsible for the mechanical transduction.

## VII. Effective frequency shift dependence as a function of intracavity power

Equation (1) from the main text has been combined with equation S4 in order to simulate the effective mechanical frequency as a function of laser driving wavelength and the corresponding mechanical spectrum for a second chiplet [19], as it can be seen from **Figure 6**.

In addition, the measured data from that same chiplet can be seen in **Figure 7**. Here, the power dependence of the effective mechanical frequency can be seen. As the intracavity power increases, so does the effective mechanical frequency shift.

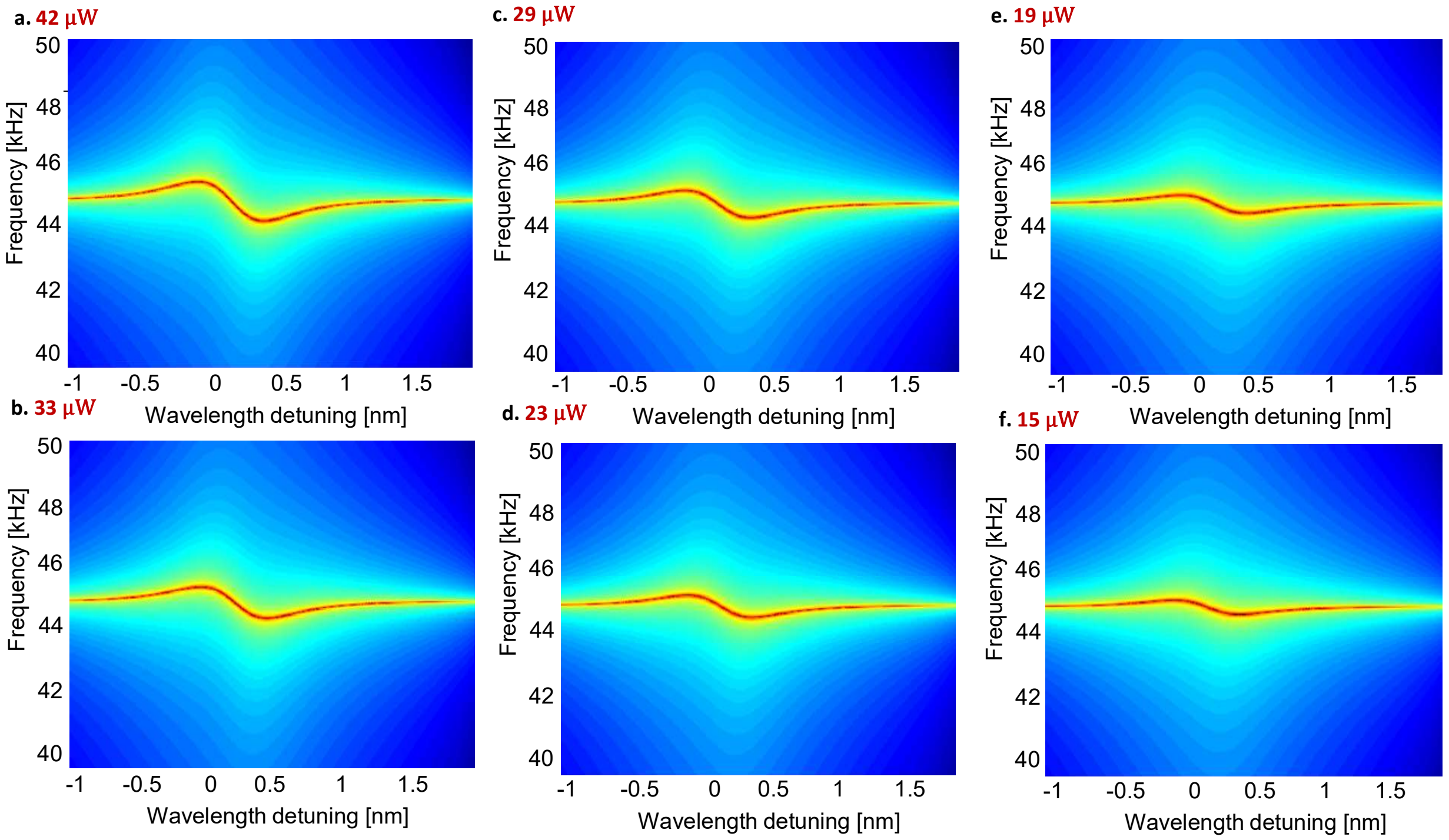

**Figure 6.- Simulated frequency displacement as a function of detunning.**

Simulated power spectrum for the optomechanical accelerometer for different intracavity powers. **a,** 42 µW. **b,** 33 µW. **c,** 29 µW. **d,** 23 µW. **e,** 19 µW. **f,** 15 µW.

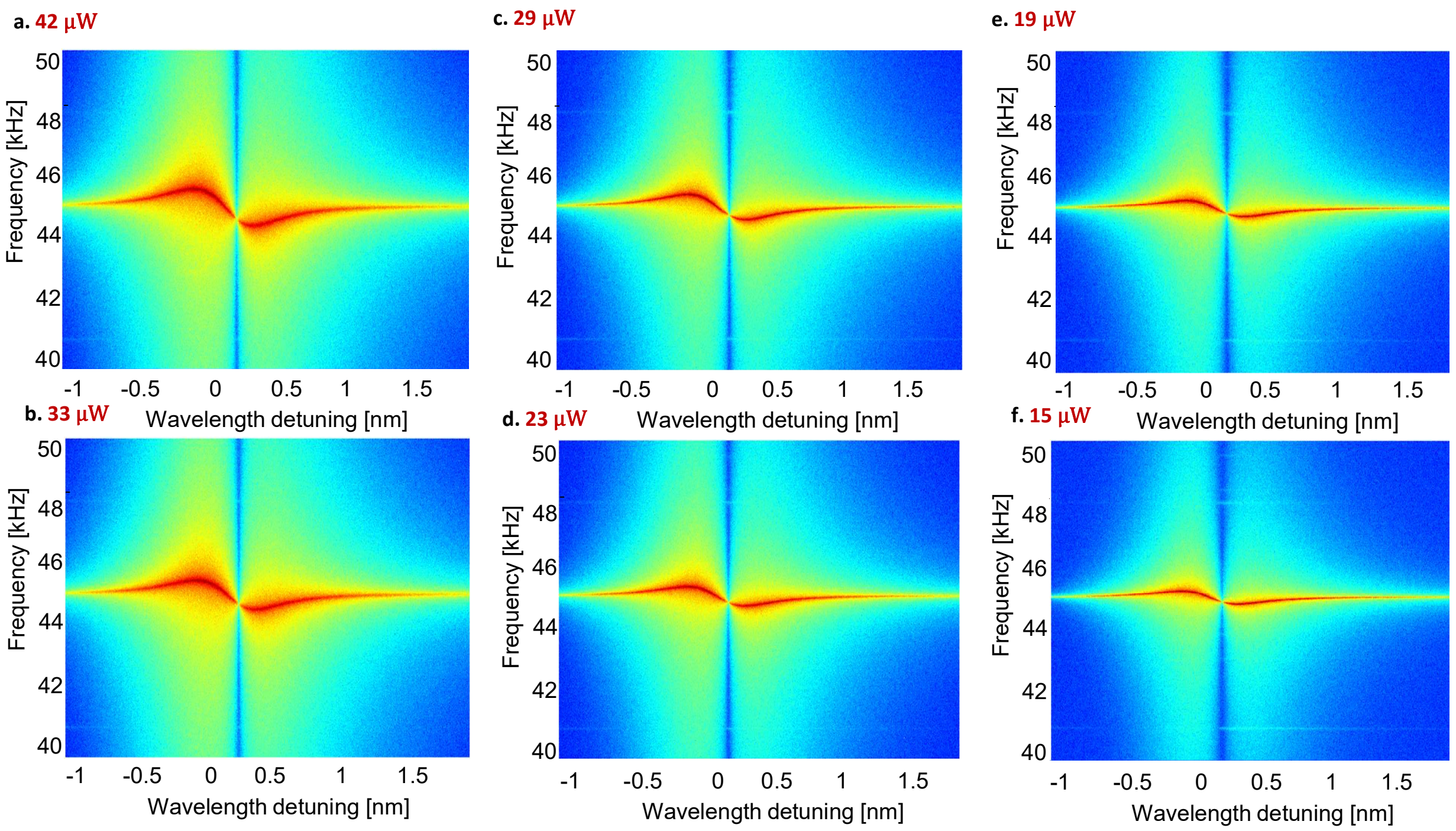

**Figure 7.- Meassured frequency displacement as a function of detunning.**

Measured power spectrum for the optomechanical accelerometer for different intracavity powers. **a,** 42 μW. **b,** 33 μW. **c,** 29 μW. **d,** 23 μW. **e,** 19 μW. **f,** 15 μW.

## VIII. Measurement setup description

The optomechanical accelerometer has been carefully aligned with the goniometer with a set of screws designed for this purpose. The three Attocube axis were placed to support the input and output optical fibers as well as the optomechanical chiplet as shown in **Figure 8a**. The sensing axis of the optomechanical transducer is labeled as $x_o$, and it has been placed along the gold thermoelectric controller. The direction of the applied acceleration ($A_c$) is marked in **Figure 8b,** where the axis intersection matched the position of the photonic crystal.

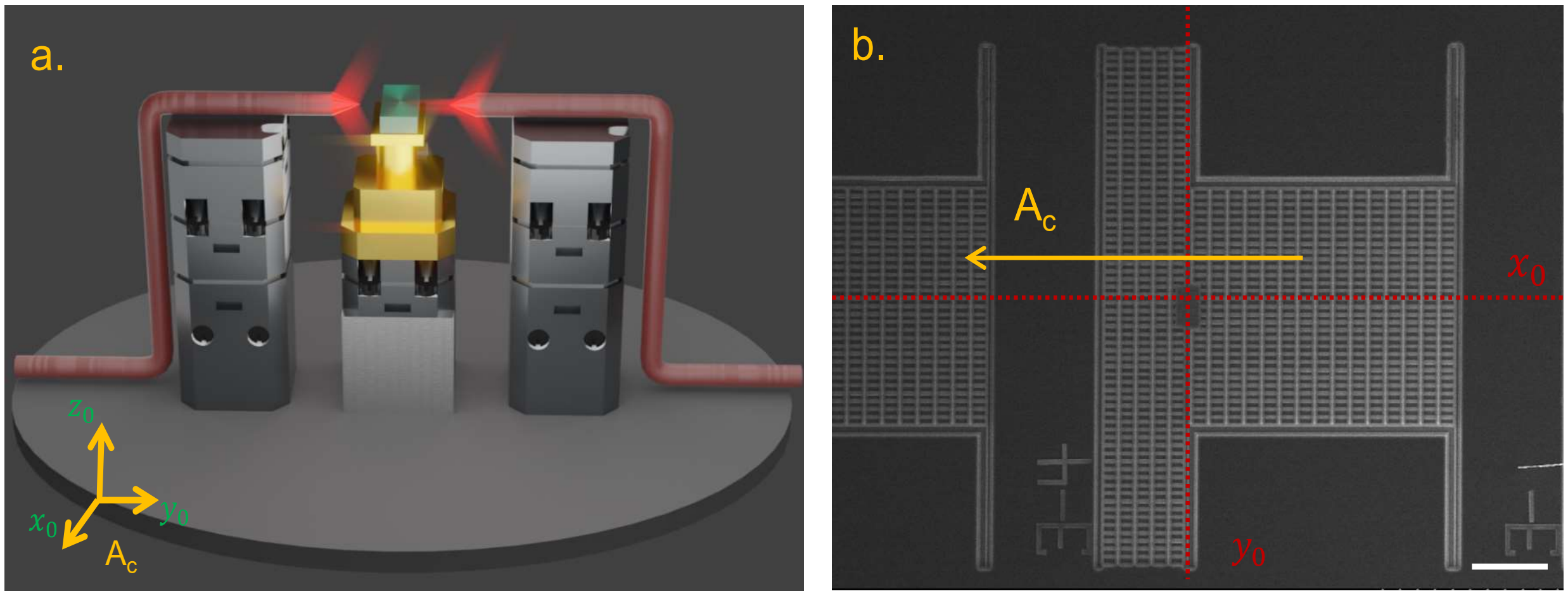

**Figure 8.- Chiplet positioning and specific force application.**

**a,** Vacuum chamber inset. Figure shows relative position of the optomechanical inertial accelerometer with respect to the internal Attocube towers and goniometers. Axis shows the relative orientation as described in the main text. $x$-axis is the sensitive axis. **b,** Optomechanical inertial accelerometer positioning yellow arrow shows the direction of the applied acceleration, which matched the $x$-axis for reference.

## IX. References

[1] B. Abbott, R. Abbott, T. D. Abbott, M. Abernathy, F. Acernese, K. Ackley and LIGO Scientific Collaboration and Virgo Collaboration, "Observation of gravitational waves from a binary black hole merger," *Physical Review Letters,* vol. 116, no. 6, 2016.

[2] Y. Ekim, J. Carballo and K. Takeda, "A survey of Autonomous driving: Common practices and emerging technologies," *IEEE Access,* vol. 8, p. 58443958469, 2020.

[3] D. Israel, K. Mauldin, C. Roberts, J. Mitchell, A. Pilkkinen, L. V. Cooper, M. Johnson, S. Christe and C. Gramling, "LunaNet: A flexible and extensible lunar exploration communications and navigation infrastructure," *2020 IEEE Aerospace Conference,* pp. 1-14, 2020.

[4] Y. Wang, H. Ding, X. Le, W. Wang and J. Xie, "A MEMS piezoelectric in-plane resonant accelerometer based on aluminum nitride with two-stage microleverage mechanism," *Sensors and Actuators A: Physical,* vol. 254, pp. 126-133, 2017.

[5] S. Shin, A. Daruwalla, M. Gong, A. Wen and F. Ayazi, "A piezoelectric resonant accelerometer for above 140db linear dynamic ranger high-G applications," *2019 20th International Conference on Solid-State Sensors, Actuators and Microsystems,* pp. 503-506, 2019.

[6] C. Wang, F. Chen, Y. Wang, S. Sadeghpour, C. Wang, M. Baijot, R. Esteves, C. Zhao, J. Bai, H. Liu and M. Kraft, "Micromachined accelerometers with sub-ug/Hz^1/2 noise floor: A review," *Sensors 20,* no. 14, p. 4054, 2020.

[7] A. Shkel, ""Precision navigation and timing enabled by microtechnology: Are we there yet?," *IEEE In Sensors,* pp. 5-9, 2010.

[8] Y. Huang, J. G. Flor Flores, Y. Li, W. Wang, D. Wang, N. Goldberg, J. Zheng, M. Yu, M. Lu, M. Kutzer, D. Rogers, D. Kwong, L. Churchill and C. W. Wong, "A Chip-Scale Oscillation-Mode Optomechanical Inertial Sensor Near the Thermodynamical Limits," *Laser & Photonics Reviews,* vol. 14, no. 5, p. 1800329, 2020.

[9] G. Anetsberger, E. Gavartin, O. Arcizet, Q. Unterreithmeier, E. M. Weig, M. L. Gorodetsky, J. P. Kotthaus and T. J. Kippenberg, "Measuring nanomechanical motion with an imprecision below the standard quantum limit," *Kippenberg,* vol. 82, no. 6, p. 061804, 2010.

[10] Y. P. Xu and et al., MEMS silicon oscillating accelerometers and readout circuits, Delft: River Publishers, 2019.

[11] Z. Yang, J. Zhao, X. Wang, G. Xia, A. P. Qiu, Y. Su and Y. P. Xu, "A sub-μg bias-instability MEMS oscillating accelerometer with an ultra-low-noise read-out circuit in CMOS," *IEEE Journal of Solid-State Circuits,* vol. 50, no. 9, pp. 2113-2126, 2015.

[12] D. Mason, J. Che, M. Rossi, Y. Tsaturyan and A. Schliesser, "Continuous force and displacement measurement below the standard quantum limit," *Nature Physics,* vol. 15, no. 8, pp. 745-749, 2019.

[13] K. Komori, Y. Enomoto, C. P. Ooi, Y. Miyazaki, N. Matsumoto, V. Sudhir, Y. Michimura and M. Ando, "Attonewton-meter torque sensing with a macroscopic optomechanical torsion pendulum," *Physical Review A,* vol. 101, no. 1, p. 011802, 2020.

[14] Y. L. Li and P. F. Barker, "Characterization and testing of a micr-g whispering gallery mode optomechanical accelerometer," *Journal of Lightwave Technology,* vol. 36, no. 18, pp. 3919-3926, 2018.

[15] A. Krause, M. Winger, T. Blasius, Q. Lin and O. Painter, "A high-resolution microchip optomechanical accelerometer," *Nature Photonics,* vol. 6, no. 11, pp. 768-772, 2012.

[16] Y. Li, J. Zheng, J. Gao, J. Shu, M. Aras and C. W. Wong, "Design of dispersive optomechanical coupling and cooling in ultrahigh-Q/V slot-type photonic crystal cavities," *Optics Express,* vol. 18, no. 23, pp. 23844-23856, 2010.

[17] K. Schneider, Y. Baumgartner, S. Hönl, P. Welter, H. Hahn, D. Wilson, L. Czornomaz and P. Seidler, "Optomechanics with one-dimensional gallium phosphide photonic crystal cavities," *Optica,* vol. 6, no. 5, pp. 577-584, 2019.

[18] M. Gorodetksy, A. Schliesser, G. Anetsberger, S. Deleglise and T. Kippenberg, "Determination of the vacuum optomechanical coupling rate using frequency noise calibration," *Optics Express,* vol. 18, no. 22, pp. 23236-23246, 2010.

[19] J. G. Flor Flores, Y. Huang, L. Li, V. Iaia, M. Tu, D.-L. Kwong and C. W. Wong, "Power-dependence of high-Q optomechanical oscillators: from pre-oscillation, to oscillation slope, to Drude-plasma," *Conference on Lasers and Electro-Optics (CLEO),* pp. 1-2, 2017.